\documentstyle[eqsecnum,pre,preprint,aps,epsfig]{revtex}
\tightenlines

\def\pr#1#2#3{ Phys. Rev. ${\bf{#1}}$ (#2) #3}
\def\prl#1#2#3{ Phys. Rev. Lett. ${\bf{#1}}$ (#2) #3}
\def\pl#1#2#3{ Phys. Lett. ${\bf{#1}}$ (#2) #3}
\def\prep#1#2#3{ Phys. Rep. ${\bf{#1}}$ (#2) #3}

\def\epjc#1#2#3{ Eur. Phys. Jour. ${\bf{#1}}$ (#2) #3}

\newcommand{\sss}{\tilde{s}_l}
\newcommand{\ccc}{\tilde{c}_l}

\newcommand{\bl}{\begin{list}}

\newcommand{\el}{\end{list}}

\newcommand{\be}{\begin{equation}}

\newcommand{\ee}{\end{equation}}
\newcommand{\bs}{\begin{mathletters}} 
\newcommand{\es}{\end{mathletters}} 

\newcommand{\baa}{\begin{eqnarray}}
\newcommand{\eaa}{\end{eqnarray}}

\newcommand{\ba}{\bs\begin{eqnarray}}
\newcommand{\ea}{\end{eqnarray}\es}

\newcommand{\bt}[1]{\bs\label{#1}\begin{eqnarray}}
\newcommand{\et}{\end{eqnarray}\es}

\newcommand{\reffig}[1]{Fig.~(\ref{#1})}

\newlength{\www}

\begin{document}
\draft

\title{New Physics Effects from $e^+e^-\to f\bar f$ at a Linear Collider:
the Role of $A_{LR, \mu}$}

\author{\underline{M.Beccaria}${}^{a, b, *}$, F. M. Renard${}^c$, S. Spagnolo${}^d$, C. Verzegnassi${}^{e,b}$}

\address{
\vspace{0.5cm}
${}^a$ 
Dipartimento di Fisica, Universita' di Lecce \\
Via Arnesano, 73100 Lecce, Italy.\\ \vspace{0.2cm}
${}^b$ 
INFN, Sezione di Lecce\\
Via Arnesano, 73100 Lecce, Italy.\\ \vspace{0.2cm}
${}^c$ 
Physique Math\'{e}matique et Th\'{e}orique, UMR 5825\\
Universit\'{e} Montpellier II,  F-34095 Montpellier Cedex 5.\\ \vspace{0.2cm}
${}^d$ 
Rutherford Appleton Laboratory - Particle Physics Department \\
Chilton, Didcot, Oxfordshire OX11 0QX. \\ \vspace{0.2cm}
${}^e$ 
Dipartimento di Fisica Teorica, Universit\`a di Trieste, \\
Strada Costiera
 14, Miramare (Trieste)
}

\maketitle

\begin{abstract}
We discuss New Physics effects in fermion pair production at LC in the
framework of the ``Z-peak subtracted approach'', a theoretical scheme that
exploits the experimental measurements at LEP1 and SLC as input
parameters.
In particular, we discuss 
the role of the longitudinal polarization asymmetry $A_{LR,\mu}$ which
turns out to be a very sensitive probe to New Physics of universal type.
The extension of the method to non universal effects is discussed and an application is given in
two examples: general contact interactions and low energy gravity models with graviton exchange.
\end{abstract}


\section{Introduction}
\label{sec:intro}

The calculation of radiative corrections (RC) at future Linear Colliders
(LC) facilities~\cite{LC} is more difficult than at LEP1 energies: processes
are no more dominated by Born diagrams and one loop corrections grow
significantly with energy. The simple scenario describing $Z$ peak
physics in terms of  dominating single $Z$
resonant exchange is not valid at a LC with energy of the order
of 1 TeV. 

In such a situation, it seems 
difficult to give a simple parametrization of RC in
the same spirit of LEP1 approaches like those leading for instance to
Peskin-Takeuchi $S$, $T$ parameters~\cite{PT} or Altarelli-Barbieri
$\varepsilon_1$, $\varepsilon_3$ ones~\cite{AB}.

However, as we briefly review in this paper, there exists a simple
scheme already exploited in LEP2 data analysis which, at least for
New Physics of universal type (UNP), provides such a simple parametrization
at LC energies too~\cite{delta}. This is the so called ``Z-peak subtracted
approach'' fully developed and illustrated in a series of dedicated
papers~\cite{Zsub} (see also~\cite{susy} for a recent application to 
supersymmetric corrections).
The basic idea of the scheme is to take LEP1 measurements as
 reference point and to describe virtual corrections at higher
energies as deviations with respect to this point.
With this aim, the conventional input parameter $G_\mu$ is replaced by 
certain observables measured at LEP or SLC on top of Z resonance. 
This approach is natural and effective: universal RC are
parametrized in terms of three subtracted quantities 
where all $q^2$ independent effects
disappear. The experimental error in the LEP1 measurements becomes
thus a source of theoretical error of the scheme.
Energy independent effects are constrained at low energy and left
completely aside  in the high energy analysis.

The plan of the paper is the following. In Sec.~(\ref{sec:zsub}) we review
some technical details on the Z peak subtracted approach. 
In Sec.~(\ref{sec:unp}) we list the expressions for UNP contributions in
two important specific models and within our Z-peak subtracted parametrization.
Sec.~(\ref{sec:nu}) is devoted to the extension of the method to 
the case of two particularly simple models of {\it non universal} type, for which it
is possible to perform an analogous analysis.
In Sec.~(\ref{sec:asym}) we discuss the special role played by the
longitudinal polarization asymmetry $A_{LR, \mu}$. 
Finally, in Sec.~(\ref{sec:conc}) we summarize our results.

\section{Review of the $Z$-peak subtracted approach}\
\label{sec:zsub}

\begin{figure}
\begin{center}
\leavevmode
\epsfig{file=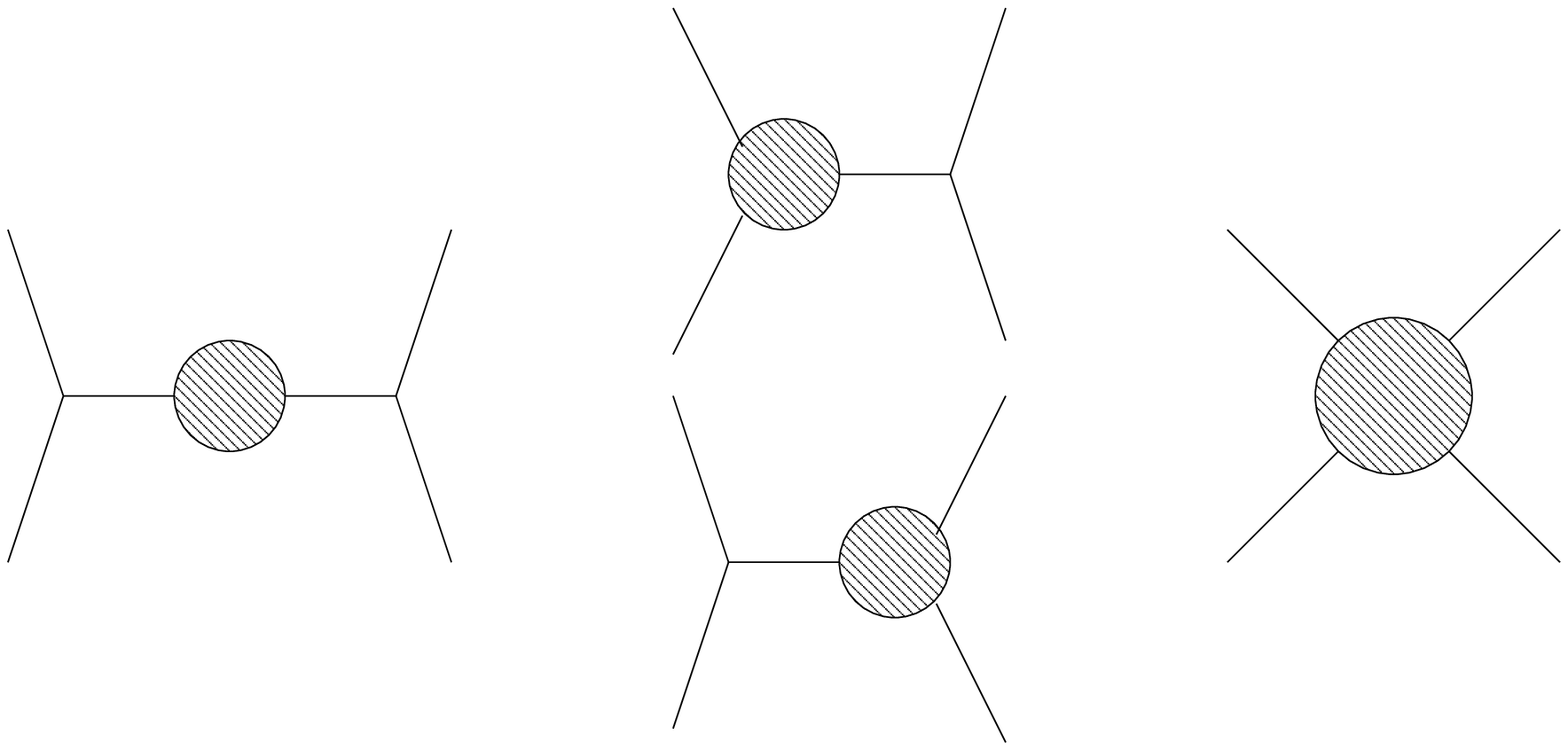,width=0.8\textwidth,angle=0}
\end{center}
\caption{One particle irreducible parts in a general $2\to 2$ process.}
\label{fig:blobs}
\end{figure}

Let us consider the process $e^+e^-\to l^+ l^-$ for the production of
a lepton pair. The analysis could be done for a generic fermion in the final
state~\cite{Zsub}, but to simplify the discussion we consider here the
simpler case of the leptonic channel. The invariant scattering amplitude
receives radiative corrections from the one particle irreducible self
energies, initial and final vertices and boxes as shown schematically
in \reffig{fig:blobs}. 
The one loop amplitude can be written in the form of a
modified Born expression  
\baa
A(s,\theta) &=& \frac{i}{q^2} v_\mu^{(\gamma)} v^{(\gamma)\mu}
(1-{\widetilde F}_\gamma) + \nonumber \\
&+&
\frac{i}{q^2-M_{0Z}^2} v_\mu^{(Z)} v^{(Z)\mu} \left(1-\frac{{\widetilde A}_Z}{q^2-M_{0Z}^2}
\right) + \\
&-& \frac{2i}{q^2-M_{0Z}^2} v_\mu^{(\gamma)} v^{(Z)\mu} \frac{{\widetilde
A_{\gamma Z}}}{q^2} \nonumber ,
\eaa
where we have introduced the photon and Z Lorentz structures
\be
v_\mu^{(\gamma)} = -|e_0| \langle l_2 |
J_\mu^{(\gamma)}(0)|l_1\rangle,\ 
J_\mu^{(\gamma)} = \sum_i Q_i \bar\psi_i\gamma_\mu\psi_i ,
\ee
\be
v_\mu^{(Z)} = -\frac{|e_0|}{2s_0c_0}  \langle l_2 | J_\mu^{(Z)}(0)|l_1\rangle,\ 
J_\mu^{(Z)} = \sum_i \bar\psi_i \gamma_\mu (g_{Vi}^0-\gamma_5
g_{Ai}^0)\psi_i ,
\ee
with $g_{Ai}^0 = I_{3L,i}$, $g_{Vi}^0 = I_{3L, i}-2Q_i s_0^2$. All the
virtual contributions are hidden into the three quantities
$\widetilde F_\gamma$, $\widetilde A_Z$ and $\widetilde A_{\gamma
Z}$ after the projection of self energies, vertices and boxes on the
proper $\gamma$ and $Z$ structures.
This decomposition can be shown to be gauge invariant~\cite{sirlin} as
a consequence of the independence of these structures.

The explicit expression of the gauge invariant combinations is 
\be
{\widetilde F}_\gamma = F_\gamma-2(\Gamma_\mu^{(\gamma)},
v_{\mu}^{(\gamma)}) - A^{Box}_{\gamma\gamma} ,
\ee
\be
\frac{{\widetilde A}_Z}{q^2-M_{0Z}^2} = \frac{A_Z}{q^2-M_{0Z}^2} -2(\Gamma_\mu^{(Z)}, v_{\mu}^{(Z)})
-A^{Box}_{ZZ} ,
\ee
\baa
\frac{{\widetilde A}_{\gamma Z}}{q^2} &=& 
\frac{A_{\gamma Z}}{q^2} - (\Gamma_\mu^{(Z)}, v_\mu^{(\gamma)}) -\frac{q^2-M_{0Z}^2}{q^2} 
(\Gamma_\mu^{(\gamma)}, v_\mu^{(Z)}) + \\
&-& (q^2-M_{0Z}^2) A^{Box}_{\gamma Z} , \nonumber
\eaa
where the quantities $F_i(q^2)$ ($i=\gamma, Z, \gamma
Z$) have been extracted from the transverse part of the corresponding
self energies $A_i(q^2)$ according to 
\be
A_i(q^2) = A_i(0)+q^2 F_i(q^2) ,
\ee
and the notation $(\Gamma, \cdot)$ stands for the projection on the
$\gamma$ and $Z$ structures.

The Z-peak subtracted prescription enters at this very point. Indeed,
one can introduce the subtracted parameters (the subscript $el$
specifies the process $e^+e^-\to l^+l^-$ under consideration):
\be
{\widetilde{\Delta} \alpha}_{el} = \mbox{Re}(\tilde F_\gamma(0)-\tilde F_\gamma(q^2)) ,
\ee
\be
\tilde I_{Z, el} = \frac{q^2}{q^2-M_Z^2} \mbox{Re}(\tilde
F_Z(q^2,\theta)-\tilde F_Z(M_Z^2,\theta)) ,
\ee
\be
R_{el} = \tilde I_{Z,el}(q^2)-\tilde I_{Z,el}(M_Z^2) ,
\ee
\be
V_{el} = \mbox{Re}(\tilde F_{\gamma Z}(q^2) - \tilde F_{\gamma Z}(M_Z^2)) ,
\ee
and show that, at the one loop level, all the observables can be
rewritten in terms of ${\widetilde{\Delta} \alpha}_{el}$, $R_{el}$ and $V_{el}$ by eliminating
$G_\mu$  and introducing LEP measured quantities as new input
parameters. To be specific, let us consider the simplest example, the
cross section for $\mu^+\mu^-$ production. In this case (neglecting
the $\gamma Z$ interference term) the following typical
expression is obtained:
\baa
\sigma_\mu &=& \frac{4}{3} \pi q^2 \left\{ (\frac{\alpha}{q^2})^2(1+2{\widetilde{\Delta} \alpha}_{e\mu}) + 
\frac{1}{(q^2-M^2_Z)^2+M_Z^2 \Gamma^2}
\left(\frac{3\Gamma_l}{M_Z}\right)^2 \times \right . \nonumber \\
&\times& \left .\left(1-2R_{e\mu} -
\frac{16(1-4\sss^2)\sss\ccc}{1+\tilde v_l}
V_{e\mu}\right) \right \} ,
\eaa
where the $Z$ widths $\Gamma$ and $\Gamma_l$ appear. The quantity
$\tilde v_l$ is defined as $\tilde v_l(M_Z^2) = 1-4 {\tilde s}_l^2(M_Z^2)$ where 
$\tilde{s}^2_l=1-\tilde{c}_l^2$ 
is the effective mixing angle directly related to the measurement of
asymmetries at the Z peak. 

A similar discussion can be done in the case of hadronic observables,
in particular the light quark-antiquark production cross sections and
asymmetries.
To extend the scheme, additional input parameters must be introduced
like the Z widths and asymmetries into hadrons.
In the end, the one loop effects turn out to be parametrized in terms of
four (flavour dependent) functions~\cite{Zsub}
\be
{\widetilde{\Delta} \alpha}_{ef},\qquad R_{ef},\qquad V_{ef}^{\gamma Z},\qquad V_{ef}^{Z\gamma} .
\ee
When New Physics effects are considered, these quantities are shifted:
\be
{\widetilde{\Delta} \alpha}_{ef} \to {\widetilde{\Delta} \alpha}_{ef} + {\widetilde{\Delta} \alpha}_{ef}^{NP} ,
\ee
and similarly for the other three. It is convenient to focus on those 
New Physics models that produce virtual contributions independent
both on the final fermion flavour and on the scattering angle
$\theta$. We denote by Universal New Physics (UNP) all such
effects. Due to universality, the simplification
$V^{\gamma Z}_{ef} = V^{Z \gamma}_{ef}$ occurs and the deviation with
respect to the Standard Model can be parametrized by three functions
of $q^2$ only (here, we are allowed to omit the $ef$ label)
\be
{\widetilde{\Delta} \alpha}^{UNP}(q^2), \qquad R^{UNP}(q^2),\qquad V^{UNP}(q^2) .
\ee
By construction, these are subtracted quantities that must vanish at
$q^2=0, M_Z^2, M_Z^2$ respectively. We choose therefore to write our 
final parametrization will
in terms of the three parameters
\be
\delta_z(q^2),\qquad \delta_s(q^2),\qquad \delta_\gamma(q^2) ,
\ee
defined by 
\be
\widetilde{\Delta}\alpha^{UNP}(q^2) = \frac{q^2}{M_Z^2}\ \delta_\gamma(q^2),\qquad
R^{UNP}(q^2) = \frac{q^2-M_Z^2}{M_Z^2}\ \delta_z(q^2),\qquad
V^{UNP}(q^2) = \frac{q^2-M_Z^2}{M_Z^2}\ \delta_s(q^2) .
\ee
In the following sections we shall fix $\sqrt{q^2} = 500\ \mbox{GeV}$
and consider specific NP models for which we derive bounds in the
three dimensional space $(\delta_z, \delta_s,\delta_\gamma)$.

\section{Models with Universal New Physics Corrections}
\label{sec:unp}

In this Section, we discuss the specific form of the virtual one loop 
contribution to $\delta_z$, $\delta_s$ and $\delta_\gamma$ for two models 
where such corrections are of universal type: models with anomalous
couplings and models of technicolor type, that is with strongly coupled resonances.

\subsection {Anomalous gauge couplings}

As a first UNP model we consider one with anomalous gauge
couplings (AGC) in 
the framework of~\cite{agc} restricting our analysis to dimensions six effective
terms in the Standard Model lagrangian with $SU(2)\times U(1)$ and
$CP$ invariance. As is well known, there are 4 operators affecting the
$WW\gamma$, $WWZ$ vertices at tree level and 5 operators that enter at the
one loop level by renormalizing the coupling of the first four. In a
general conventional and model independent analysis one must thus
determine four parameters. However, as shown in~\cite{agczsub}, only two parameters ($f_{DW}$
and $f_{DB}$) survive in the Z-peak subtracted approach making the
analysis and the fit to experimental data much more simple. 

The explicit expression of the UNP contribution to
$\delta_z$, $\delta_s$ and $\delta_\gamma$ are 
\be
\delta_z = 8\pi\alpha \frac{M_Z^2}{\Lambda^2} \left(\frac{\ccc^2}{\sss^2} f_{DW} + \frac{\sss^2}{\ccc^2} 
f_{DB}\right) ,
\ee
\be
\delta_s = 8\pi\alpha \frac{M_Z^2}{\Lambda^2} \left(\frac \ccc \sss f_{DW} -
\frac \sss \ccc f_{DB}\right) ,
\ee
\be
\delta_\gamma =
-8\pi\alpha\frac{M_Z^2}{\Lambda^2}\left(f_{DW}+f_{DB}\right) ,
\ee
Since we have three parameters and only two couplings, we obtain a linear constraint:
\be
\delta_z -\frac{1-2\sss^2}{\sss\ccc} \delta_s+\delta_\gamma = 0 .
\ee

\subsection{Models of Technicolor type}

Another interesting class of UNP models where the Z-peak subtracted
approach turns out to be useful is described and analyzed
in~\cite{tc}. In the Z-peak approach the correction coming from self
energies is subtracted and can be represented by a dispersion
relation. One can add the effect of possible strongly coupled
resonances by adding phenomenologically sensible contributions to the
spectral weight of the representation. The typical UNP parameters are
then the resonance mass, width and coupling $F$. In the simplest case, 
one considers just a pair of heavy vector and axial resonances with
masses much larger than $M_Z$ and $\sqrt{q^2}$ and in the zero width limit. In this
case the UNP contribution can be shown to be a function of the two
ratios $F_A/M_A$ and $F_V/M_V$ where $F_{A, V}$ and $M_{A, V}$ are the
couplings and the masses of  the the axial and vector resonances.

The contribution to $\delta_z$, $\delta_s$ and $\delta_\gamma$ are
\be
\delta_z = M_Z^2 \frac{\pi\alpha }{\sss^2\ccc^2} \left(
(1-2\sss^2)^2 \frac{F_V^2}{M_V^4}+\frac{F_A^2}{M_A^4}\right) ,
\ee
\be
\delta_s = M_Z^2 \frac{2\pi\alpha}{\sss\ccc} (1-2\sss^2) \frac{F_V^2}{M_V^4} ,
\ee
\be
\delta_\gamma = -4 \pi\alpha M_Z^2 \frac{F_V^2}{M_V^4} .
\ee
Again, we have a linear constraint in the $(\delta_z, \delta_s,
\delta_\gamma)$ space:
\be
\delta_s = -\frac{1-2\sss^2}{2\sss\ccc} \delta_\gamma .
\ee

\section{Models with Non Universal New Physics Corrections}
\label{sec:nu}

In the previous analysis we only considered
models that are {\it both} $\theta$ independent {\it and} of universal 
"smooth" type, thus achieving remarkable simplifications, particularly in
our $Z$ peak subtracted approach where we have been able to reduce the
number of parameters to the triplet $\delta_\gamma$, $\delta_s$ and $\delta_z$.

On the other hand, there exist also interesting models of new physics that do not meet both
previous requests. In this
final part, we have in fact extended our analysis to the treatment
of two models, that we list here following the order in which they
violate our two simplicity conditions.\par

\vskip 0.5cm
\noindent
{\bf a) Contact interactions}
\vskip 0.5cm

\noindent
The following interaction
\be
{\cal L}={G\over\Lambda^2}\bar \Psi\gamma^{\mu}(a_e-b_e\gamma^5)\Psi
\bar\Psi\gamma_{\mu}(a_f-b_f\gamma^5)\Psi
\label{cont}\ee
\noindent
was first introduced with the idea of
compositeness \cite{contact}, but it applies to any 
virtual NP effect (for example higher vector boson exchanges)
satisfying chirality conservation 
(Vector and Axial Lorentz structures) and whose effective scale
$\Lambda$ is high enough so that one can restrict to six dimensional 
operators.\par
The parameters $a_e,b_e,a_f,b_f$ can be
adjusted in order to describe all kind of chiral couplings. For each
choice of pair of chiralities among 
L($a=b=1/2$), R($a=-b=1/2$), V($a=1,~b=0$),
A($a=0,~b=1$), there is only one free
parameter.

These models 
are not of universal type, but retain the property of being $\theta$
independent. Their contribution to observables can be easily "projected"
on $\delta_i$ leading to:

\baa
&&\delta_{z,ef}=-({GM^2_Z\over\Lambda^2}){4\tilde s^2_l\tilde
c^2_lb_eb_f
\over e^2I_{3e}I_{3f}}\nonumber\\
&&\delta^{\gamma Z}_{s,ef}=-({GM^2_Z\over\Lambda^2})
{4\tilde s_l\tilde c_l(a_e-b_e\tilde v_l)b_f
\over e^2Q_eI_{3f}}\nonumber\\
&&\delta^{Z\gamma}_{s,ef}=-({GM^2_Z\over\Lambda^2})
{4\tilde s_l\tilde c_l(a_f-b_f\tilde v_f)b_e
\over e^2Q_fI_{3e}}\nonumber\\
&&\delta_{\gamma,ef}=({GM^2_Z\over\Lambda^2})
{(a_e-b_e\tilde v_l)(a_f-b_f\tilde v_f)
\over e^2Q_eQ_f}
\eaa

\vskip 0.5cm
\noindent
{\bf a) Manifestations of extra dimensions}
\vskip 0.5cm

\noindent
Finally we consider a model for which neither universality 
nor $\theta$ independence are retained. Recently, an intense activity
has been developed on possible low energy effects of graviton exchange. 
The following matrix element for the 
4-fermion process $e^+e^-\to \bar{f} f$ is predicted~\cite{ed}:
\be
{\lambda\over\Lambda^4}[\bar e\gamma^{\mu}e\bar
f\gamma_{\mu}f(p_2-p_1).(p_4-p_3)-\bar e\gamma^{\mu}e\bar f\gamma^{\nu}f
(p_2-p_1)_{\nu}(p_4-p_3)_{\mu}]
\ee
leading to:
\baa
&&\delta_{z,ef}=-({\lambda M^2_Z q^2\over \Lambda^4})
{4\tilde s^2_l\tilde c^2_l
\over e^2I_{3e}I_{3f}}\nonumber\\
&&\delta^{\gamma Z}_{s,ef}=({\lambda M^2_Z q^2\over\Lambda^4})
{2\tilde s_l\tilde c_l\tilde v_l
\over e^2Q_eI_{3f}}\nonumber\\
&&\delta^{Z\gamma}_{z,ef}=({\lambda M^2_Z q^2\over\Lambda^4})
{2\tilde s_l\tilde c_l\tilde v_f
\over e^2Q_fI_{3e}}\nonumber\\
&&\delta_{\gamma,ef}=({\lambda M^2_Z q^2\over\Lambda^4})
{(\tilde v_l\tilde v_f-2cos\theta)
\over e^2Q_eQ_f}
\eaa

\section{The specific role of $A_{LR, \mu}$}
\label{sec:asym}

The longitudinal polarization asymmetry $A_{LR}$ is a very important
observable in the physics programme of LC. In our scheme it is a very
peculiar probe to NP effects with large contributions to $\delta_s$
which, roughly speaking, is the parameter related to the virtual
corrections to the weak mixing angle.
As pointed out very clearly in~\cite{asymzsub}, the expression of the
NP corrections to $A_{LR}$ in the Z-peak subtracted scheme is quite
simple and inspiring. In the specific case of $\mu$
production\footnote{The theoretical properties of the asymmetries for
hadron production are similar, but the precision of their
current measurement and the aimed high luminosity of the LC do not encourage
their use within the Z-peak approximation.} one has
\baa
\frac{\delta A_{LR, \mu}}{A_{LR, \mu}} &=& (\Delta\alpha+R)\left(\frac{\kappa (q^2-M_Z^2)}{\kappa (q^2-M_Z^2) + q^2} + \right. \\
&-& \left .\frac{2\kappa^2 (q^2-M_Z^2)^2}{\kappa^2 (q^2-M_Z^2)^2 +
q^4}\right) - 
\frac{4\sss\ccc}{\tilde v_l} V , \nonumber
\eaa
where $\kappa = \alpha(0) M_Z/(3\Gamma_l) \simeq 2.63$.
Since $\tilde v_e = 0.074$, the last term turns out to be accidentally quite
large explaining the large sensitivity of $A_{LR, \mu}$ to radiative
corrections affecting $V$.

Repeating the analysis in the case of $\sigma_\mu$, $A_{FB, \mu}$ and
$\sigma_5$ (the cross section for the production of the five light
quarks and antiquarks) one obtains the following
numerical values at the energy $\sqrt{q^2} = 500\ \mbox{GeV}$:
\be
\frac{\delta\sigma_\mu}{\sigma_\mu} = -7.84\ \delta_z -6.82\
\delta_s+52.2\ \delta_\gamma ,
\ee
\be
\frac{\delta A_{FB, \mu}}{A_{FB, \mu}} = -21.3\ \delta_z+1.17\
\delta_s-22\ \delta_\gamma ,
\ee
\be
\frac{\delta A_{LR, \mu}}{A_{LR, \mu}} = -26.0\ \delta_z-664\
\delta_s-26.9\ \delta_\gamma ,
\ee
\be
\frac{\delta \sigma_5}{\sigma_5} = -27\ \delta_z-37.3\ \delta_s+32.1\
\delta_\gamma .
\ee
The conclusions that can be derived from these numbers are the
following: 
(a) $\sigma_\mu$ is the natural choice for probing UNP effects modifying mainly
$\delta_\gamma$, 
(b) the left-right asymmetry $A_{LR}$ is very sensitive to
$\delta_s$ as expected, 
(c) the cross section for the production of the five light quark 
depends on the three $\delta$ parameter with comparable weights and 
its inclusion in a fitting procedure provides a constraint in an independent
direction in $\delta$ space. 
The forward backward asymmetry is expected to play a
minor role in this analysis.

To be more quantitative, we consider a LC with c.m. energy
$\sqrt{q^2}=500\ \mbox{GeV}$ and luminosity ${\cal L} = 500
\ \mbox{fb}^{-1}$. We assume no deviations with respect to the Standard
Model and consider a purely statistical error on all the observables. 
We then derive bounds on the three parameters $\delta_z$, $\delta_s$
and $\delta_\gamma$ by a $\chi^2$ study based on the use of the observables
$\sigma_\mu$, $A_{FB, \mu}$ and $\sigma_5$ and 
discussing the expected improvements when $A_{LR, \mu}$ is included in
the analysis. 

\begin{figure}
\begin{center}
\leavevmode
\epsfig{file=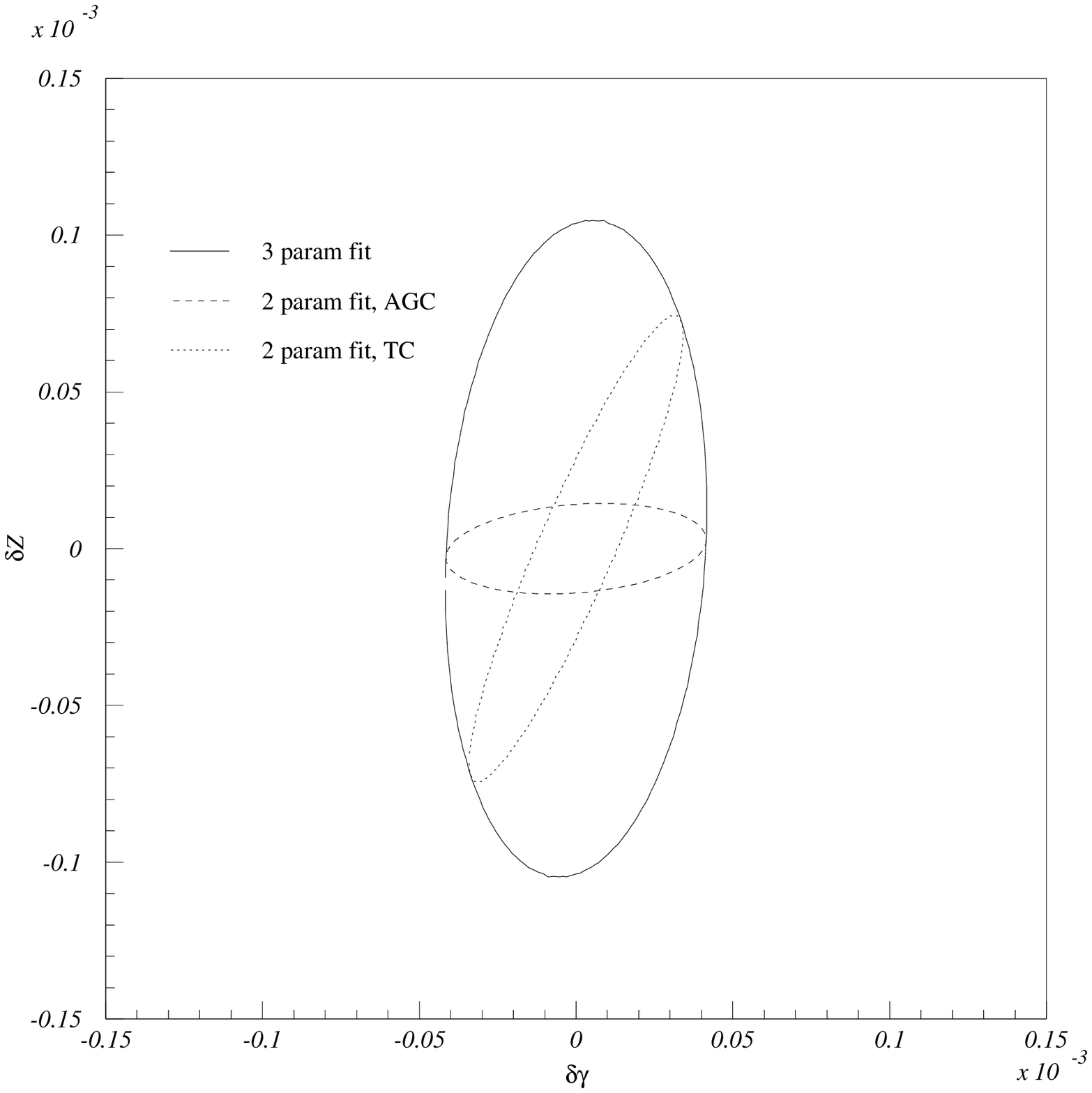,width=1.0\textwidth,angle=0}
\end{center}
\caption{Allowed domain in the $(\delta_z, \delta_\gamma)$ plane, without
$A_{LR, \mu}$.}
\label{fig:wo13}
\end{figure}

\begin{figure}
\begin{center}
\leavevmode
\epsfig{file=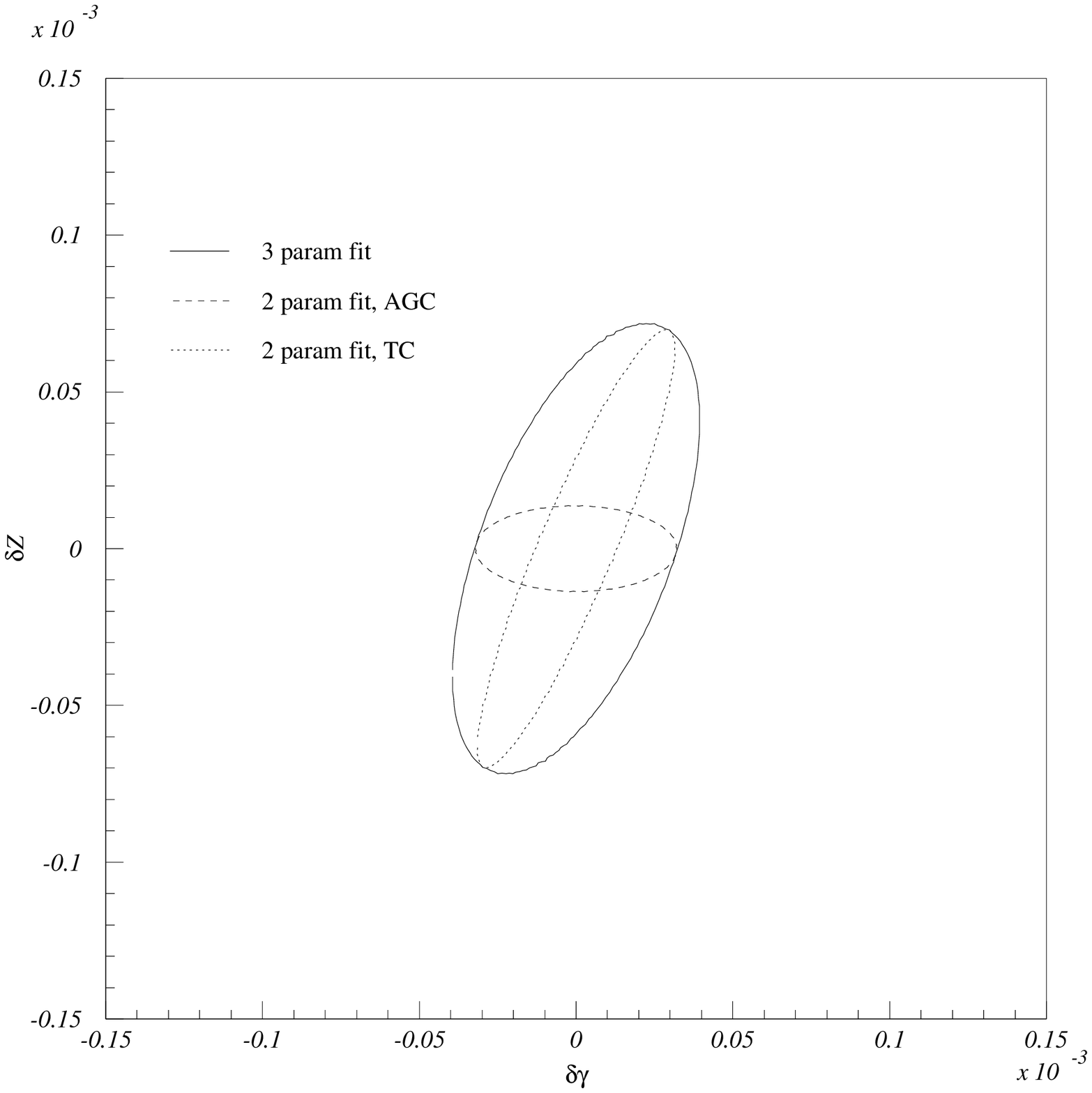,width=1.0\textwidth,angle=0}
\end{center}
\caption{Allowed domain in the $(\delta_z, \delta_\gamma)$ plane, with
$A_{LR, \mu}$.}
\label{fig:w13}
\end{figure}

\begin{figure}
\begin{center}
\leavevmode
\epsfig{file=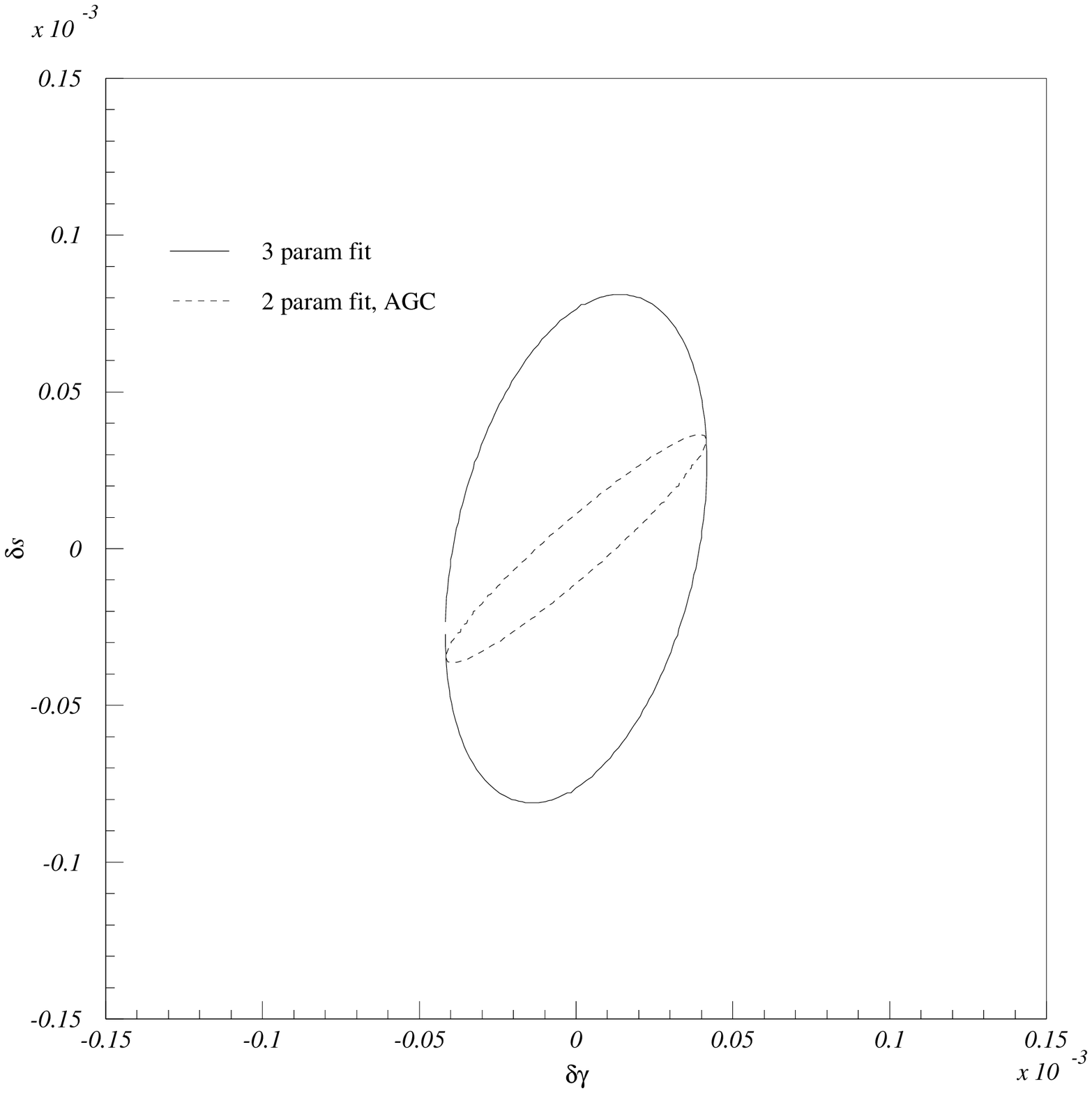,width=1.0\textwidth,angle=0}
\end{center}
\caption{Allowed domain in the $(\delta_s, \delta_\gamma)$ plane, without
$A_{LR, \mu}$.}
\label{fig:wo23}
\end{figure}

\begin{figure}
\begin{center}
\leavevmode
\epsfig{file=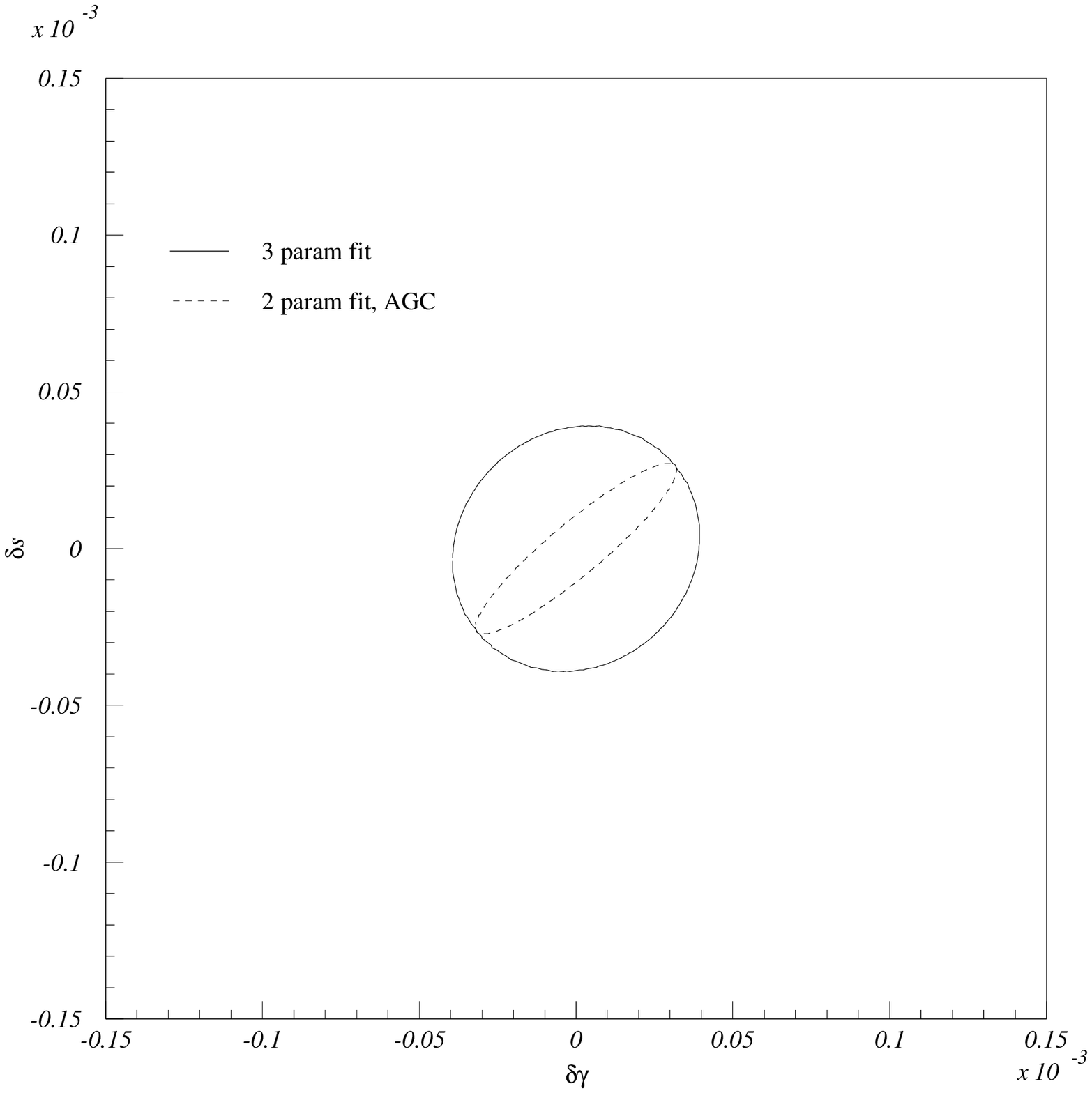,width=1.0\textwidth,angle=0}
\end{center}
\caption{Allowed domain in the $(\delta_s, \delta_\gamma)$ plane, with
$A_{LR, \mu}$.}
\label{fig:w23}
\end{figure}

\begin{figure}
\begin{center}
\leavevmode
\epsfig{file=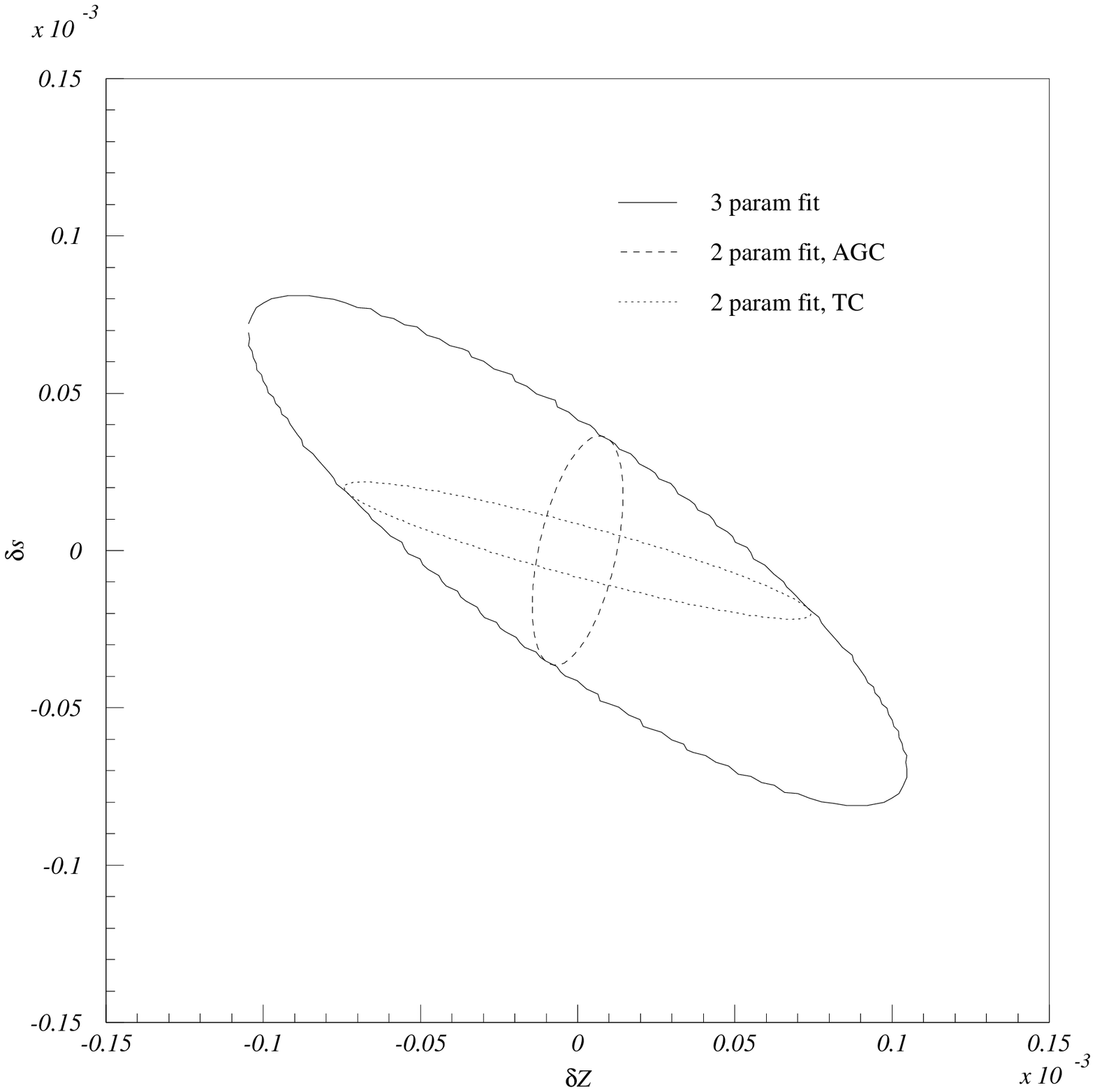,width=1.0\textwidth,angle=0}
\end{center}
\caption{Allowed domain in the $(\delta_z, \delta_s)$ plane, without
$A_{LR, \mu}$.}
\label{fig:wo12}
\end{figure}

\begin{figure}
\begin{center}
\leavevmode
\epsfig{file=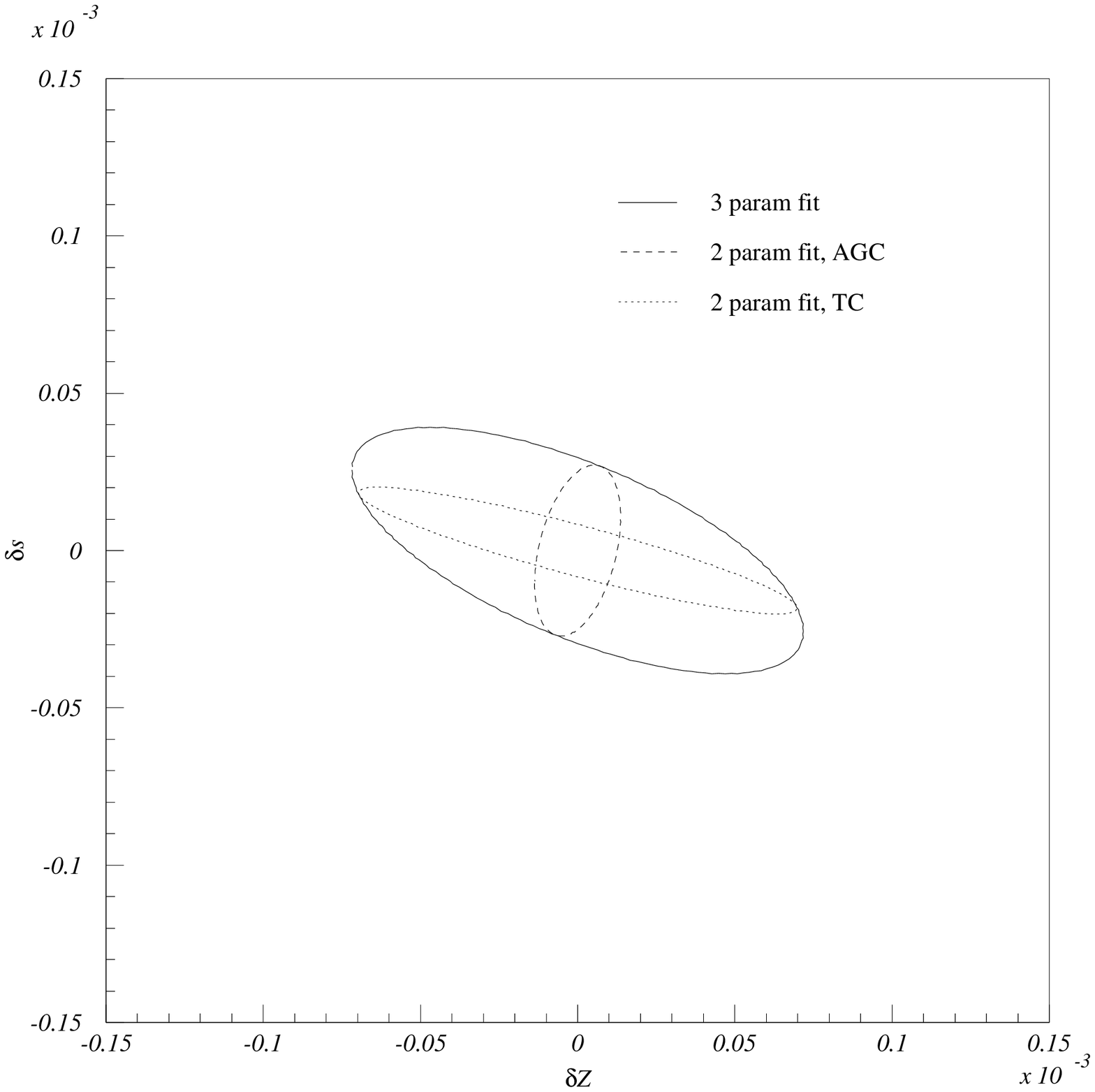,width=1.0\textwidth,angle=0}
\end{center}
\caption{Allowed domain in the $(\delta_z, \delta_s)$ plane, with
$A_{LR, \mu}$.}
\label{fig:w12}
\end{figure}

In Figs.~(\ref{fig:wo13},\ref{fig:w13}) we show the allowed
1$\sigma$ region projected onto the $(\delta_z, \delta_\gamma)$ plane
without and with the asymmetry. Both for AGC and TC we also show the projection of the two
ellipses resulting from the intersection of the three dimensional
ellipse with the plane representing the linear constraints of the
specific considered NP. 
%
%
Figs.~(\ref{fig:wo23},\ref{fig:w23}) and
Figs.~(\ref{fig:wo12},\ref{fig:w12}) are similar,
but in the $(\delta_s, \delta_\gamma)$ and $(\delta_z, \delta_s)$ planes.

\begin{figure}
\begin{center}
\leavevmode
\epsfig{file=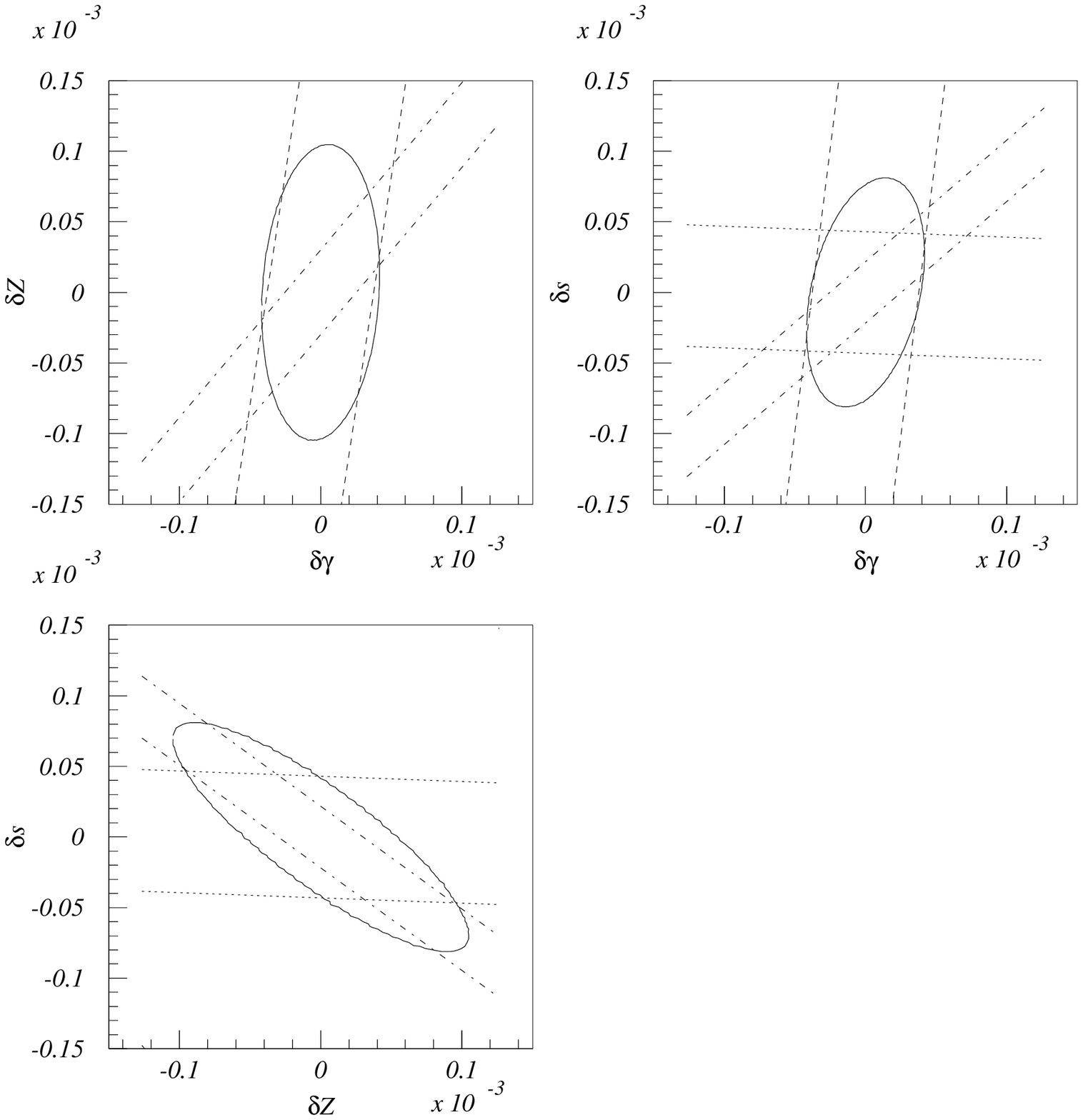,width=1.0\textwidth,angle=0}
\end{center}
\caption{Projection of the three dimensional ellipse in $(\delta_z,
\delta_s, \delta_\gamma)$ space onto the three coordinate planes. Without 
$A_{LR, \mu}$. The dashed, dot-dashed and dotted stripes are the
independent $1\sigma$ regions for $\sigma_\mu$, $\sigma_5$ and $A_{LR,
\mu}$ respectively.}
\label{fig:pwo}
\end{figure}

\begin{figure}
\begin{center}
\leavevmode
\epsfig{file=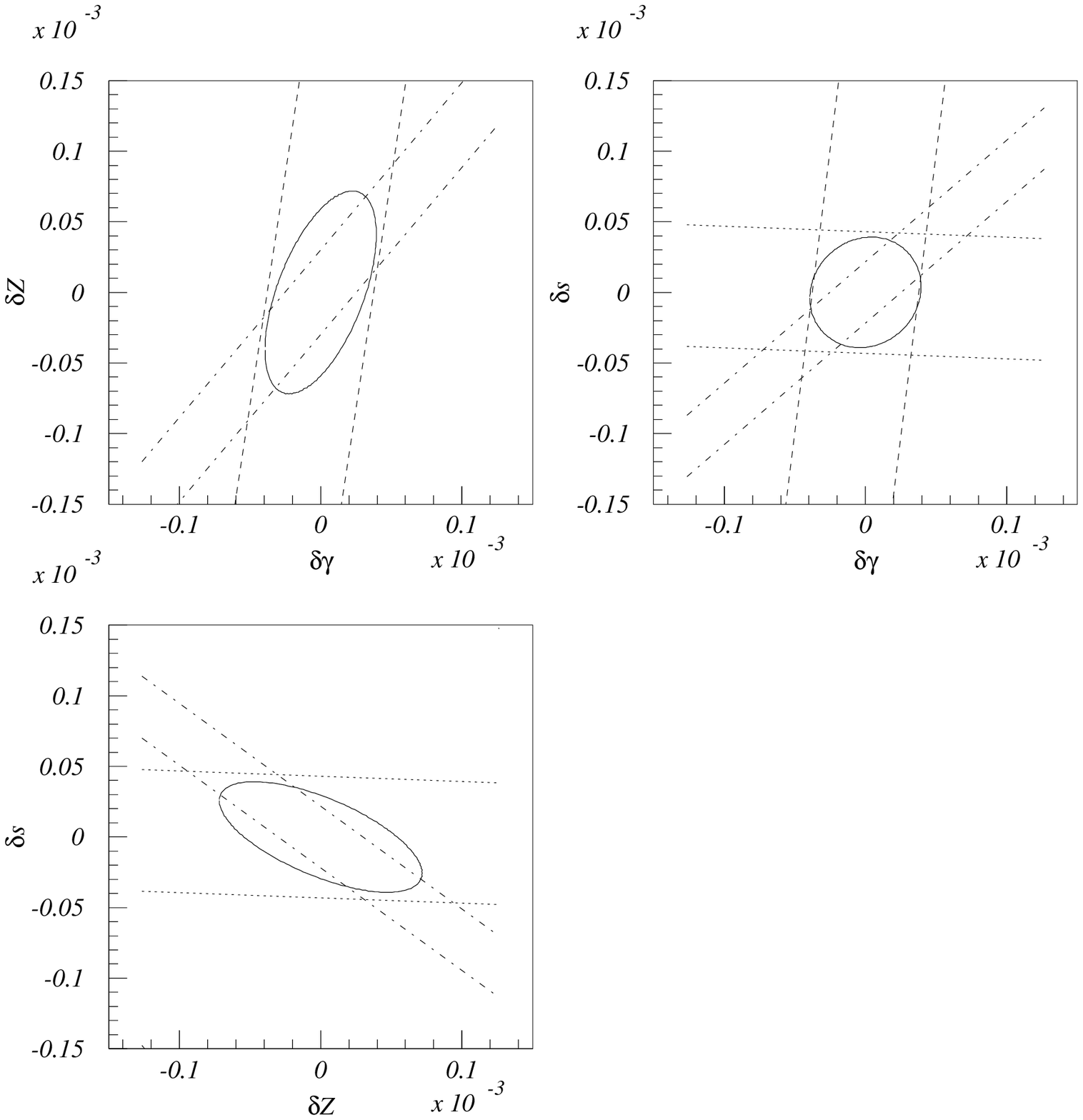,width=1.0\textwidth,angle=0}
\end{center}
\caption{Projection of the three dimensional ellipse in $(\delta_z,
\delta_s, \delta_\gamma)$ space onto the three coordinate planes. With
$A_{LR, \mu}$. The dashed, dot-dashed and dotted stripes are the
independent $1\sigma$ regions for $\sigma_\mu$, $\sigma_5$ and $A_{LR,
\mu}$ respectively.}
\label{fig:pw}
\end{figure}

A study of these pictures shows the expected general trend at
least at the level of the unconstrained three parameter fit. The
introduction of $A_{LR, \mu}$ strongly improve the bounds on the
$\delta_s$ parameter. Typically, the photon exchange parameter
$\delta_\gamma$ is the one which is less affected. To understand better
this behaviour we show in
Figs.~(\ref{fig:pwo},\ref{fig:pw}) the projection of the unconstrained three
dimensional ellipse onto the three coordinate planes. Moreover, 
we also show stripes corresponding to independent $1\sigma$ regions for
$\sigma_\mu$, $\sigma_5$ and the left-right asymmetry $A_{LR, \mu}$ 
(dashed, dot-dashed and dotted lines respectively). In other words, 
in each plane corresponding to two $\delta$ parameters, 
we set the third to zero and determine the region where the deviation 
$\delta{\cal O}$ on the observable ${\cal O}$
is smaller than the experimental error. This gives a 
rough idea of the size of each individual contribution to the bounds.
By comparing the
ellipses with and without the inclusion of the left right asymmetry
one sees that the allowed regions are essentially controlled by
$\sigma_\mu$ in the $\delta_\gamma$ direction, by $A_{LR, \mu}$ in the
$\delta_s$ direction and by $\sigma_5$ in the $\delta_z$ direction to
a somewhat smaller extent.

\begin{figure}
\begin{center}
\leavevmode
\epsfig{file=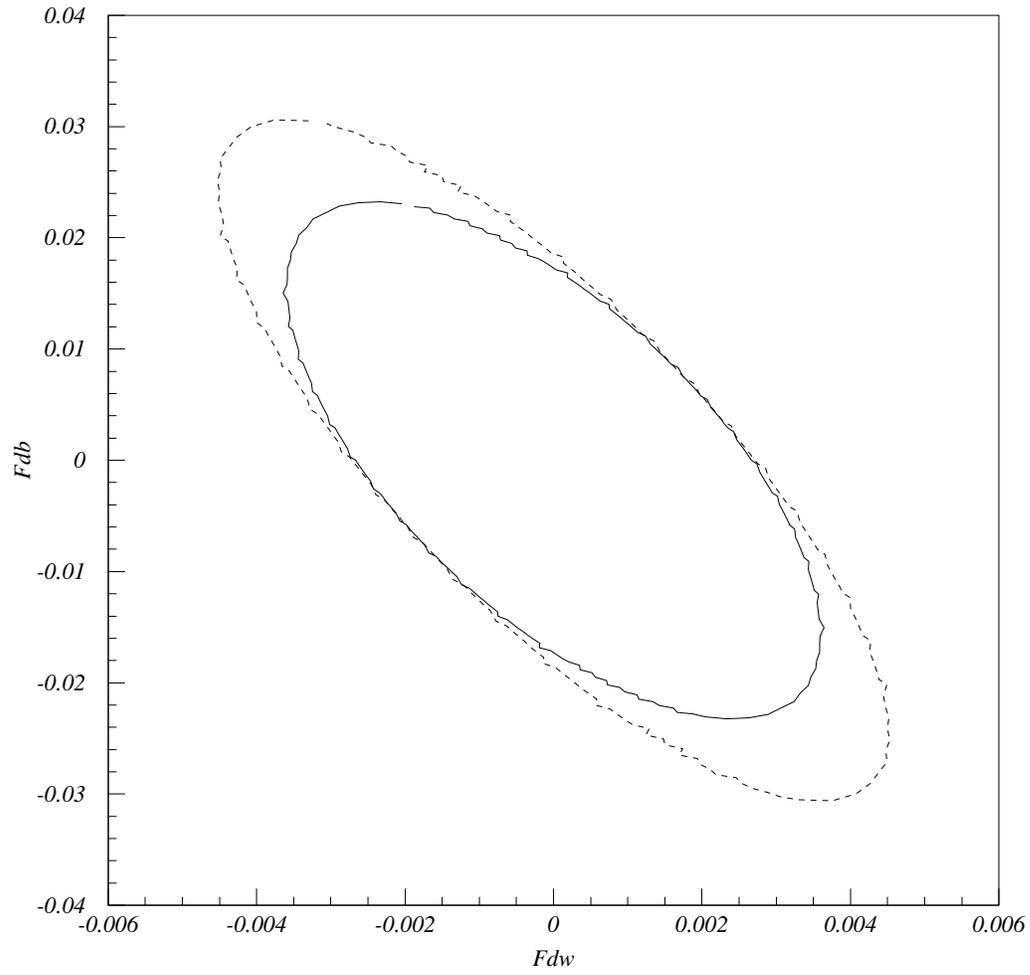,width=0.9\textwidth,angle=0}
\end{center}
\caption{Allowed regions in the $(f_{DW}, f_{DB})$ plane. With and
without $A_{LR, \mu}$.}
\label{fig:agc}
\end{figure}

\begin{figure}
\begin{center}
\leavevmode
\epsfig{file=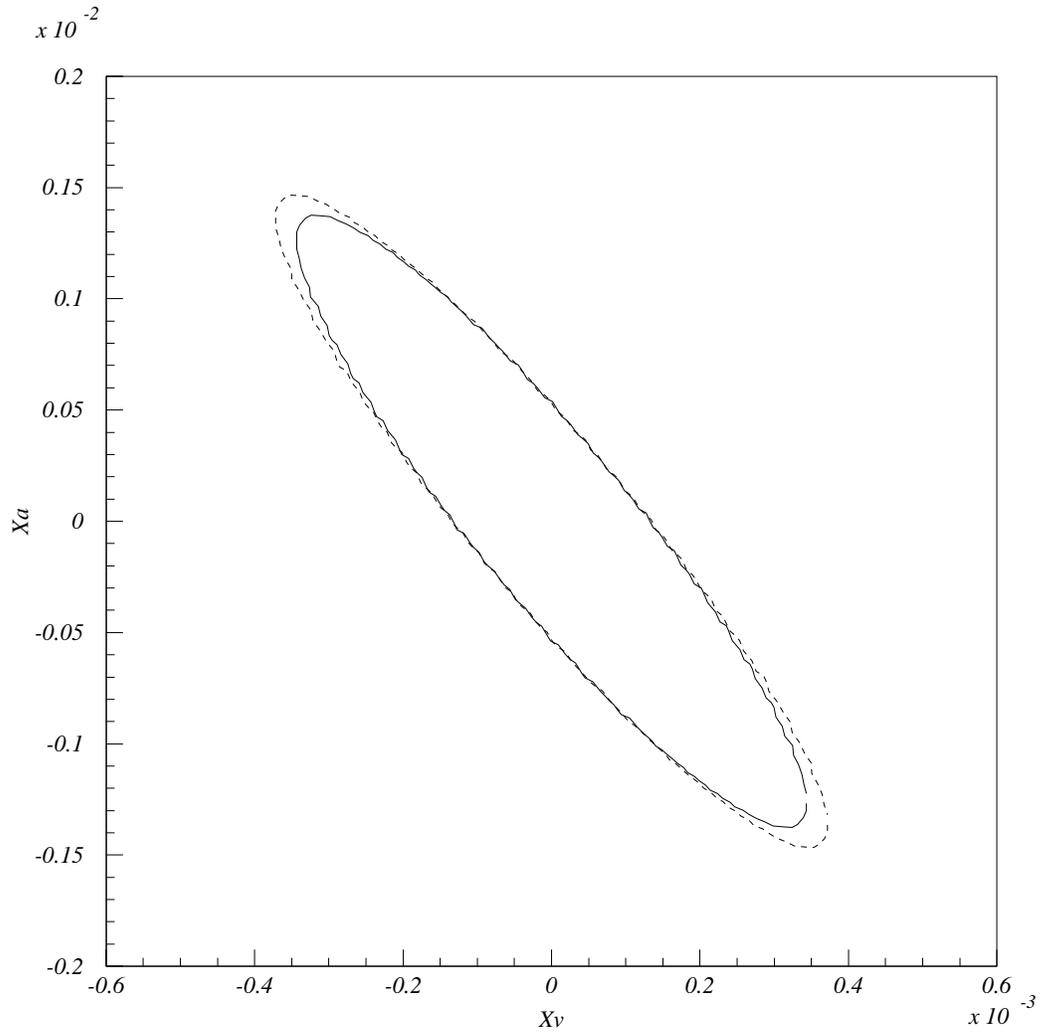,width=0.9\textwidth,angle=0}
\end{center}
\caption{Allowed regions in the $(x_V, x_A)$ plane; $x_i =
F_i^2/M_i^4$. With and without $A_{LR, \mu}$.}
\label{fig:tc}
\end{figure}

To draw conclusions and compare with similar bounds from LEP analysis,
it is convenient to consider also more physical parametrizations. 
Therefore, in Figs.~(\ref{fig:agc},\ref{fig:tc}) we turn from the
$\delta$ parametrization back to the physical couplings and plot the
allowed regions in the space of anomalous operator couplings $(f_{DB},
f_{DW})$ and of the $x$ parameters $(x_V, x_A)$ with $x_i = F_i^2/M_i^4$, with $i=V, A$.

To conclude, let us summarize the numerical values of the bounds that
we have obtained with and without the left right asymmetry in all the
considered cases (in round brackets we show the relative variation of
each parameter).

\underline{Unconstrained 3 dimensional fit:}

\begin{center}
\begin{tabular}{ccccccc}
$A_{LR}$ & $\delta_\gamma$ && $\delta_z$ && $\delta_s$ & $(10^{-4})$\\
N  & 0.41 && 1.0 && 0.82 &\\
Y  & 0.38 & (8\% ) & 0.73 & ({\bf 30\%} ) & 0.38 & ({\bf 54\%} )  
\end{tabular}
\end{center}

\underline{Anomalous gauge couplings:}

\begin{center}
\begin{tabular}{ccccccc}
$A_{LR}$ & $\delta_\gamma$ && $\delta_z$ && $\delta_s$ & $(10^{-4})$\\
N  & 0.41 && 0.15 && 0.35 &\\
Y  & 0.32 & ({\bf 23\%} ) & 0.14 & (4\%) & 0.27 & ({\bf 22\%} )  
\end{tabular}
\end{center}

\underline{Technicolor type models:}

\begin{center}
\begin{tabular}{ccccccc}
$A_{LR}$ & $\delta_\gamma$ && $\delta_z$ && $\delta_s$ & $(10^{-4})$\\
N  & 0.35 && 0.76 && 0.22 &\\
Y  & 0.32 & (9\% ) & 0.70 & (8\%) & 0.20 & (7\% )  
\end{tabular}
\end{center}

%
%
%

In terms of the physical parameters $f_{DB}$, $f_{DW}$, $x_A$, $x_V$
one obtains:

\underline{Anomalous gauge couplings:}

\begin{center}
\begin{tabular}{ccccc}
$A_{LR}$ & $f_{DW}$ &         & $f_{DB}$ \\
N        & 0.0045   &         & 0.031    \\      
Y        & 0.0036   & (20\%)  & 0.023 & (8\%) 
\end{tabular}
\end{center}

\underline{Technicolor type models:}

\begin{center}
\begin{tabular}{ccccc}
$A_{LR}$ & $x_V (10^{-3})$ &         & $x_A (10^{-3})$ \\
N        & 0.37          &         & 1.5    \\      
Y        & 0.34          & (8\%)  & 1.4 & (7\%) 
\end{tabular}
\end{center}

In the case of models of non universal type, 
the relative numerical effect of the New Physics couplings to the 
observables at $500\ \mbox{GeV}$
turns out to be the following. For Contact Interactions:
$$
\frac{\delta\sigma_\mu}{\sigma_\mu} =  (
4.51\, a_e a_f + 0.125\,b_e a_f + 
  0.125\,a_e b_f + 1.66\,b_e b_f)\frac{G}{\Lambda^2} 
$$
$$
\frac{\delta\sigma_5}{\sigma_5} =(-0.699\,a_e a_f + 1.11\,b_e a_f + 
  0.0527\,a_e b_f + 0.829\,b_e b_f)  \frac{G}{\Lambda^2}
$$
$$
\frac{\delta A_{FB,\mu}}{A_{FB, \mu}} = (- 1.96\,a_e a_f + 0.0665\,b_e a_f + 
  0.0665\,a_e b_f + 5.38\,b_e b_f) \frac{G}{\Lambda^2}
$$
$$
\frac{\delta A_{LR, \mu}}{A_{LR, \mu}} = (- 2.63\,a_e a_f + 69.8\,b_e a_f +
  25.4\,a_e b_f + 0.229\,b_e b_f ) \frac{G}{\Lambda^2}
$$
and for Extra Dimensions:
$$
\frac{\delta\sigma_\mu}{\sigma_\mu} =  2.81\cdot 10^{-3}\,\frac{\lambda}{\Lambda^4}
$$
$$
\frac{\delta\sigma_5}{\sigma_5} = - 2.82\cdot 10^{-3}\,\frac{\lambda}{\Lambda^4}
$$
$$
\frac{\delta A_{FB,\mu}}{A_{FB, \mu}} = - 0.865\,\frac{\lambda}{\Lambda^4}
$$
$$
\frac{\delta A_{LR, \mu}}{A_{LR, \mu}} =  4.19\cdot 10^{-4}\,\frac{\lambda}{\Lambda^4}
$$
where $G$ and $\lambda$ are adimensional and the New Physics mass
scales $\Lambda$ are expressed in TeV.

We can derive 68\% C.L. 
bounds with and without $A_{LR,\mu}$ for $\Lambda_{CT, ED}$ assuming $G=\lambda=1$. 
In the case of Contact Interactions, we obtain the following bounds for several cases (LL, RR, VV and AA):

\begin{center}
\begin{tabular}{|c|ccc|}
\hline\hline
$\Lambda_{CT}$	& no $A_{LR}$	& with $A_{LR}$ &\\
\hline
LL	&	59	&	67& (14 \%) \\
RR	&	56	&	66& (18 \%) \\
VV	&	48	&	48&\\
AA	&	42	&	42&\\
\hline\hline
\end{tabular}
\end{center}

In the second case of Extra Dimensions there is almost
no sensitivity to $A_{LR, \mu}$:

\begin{center}
\begin{tabular}{|c|cc|}
\hline\hline
$\Lambda_{ED}$	& no $A_{LR}$	& with $A_{LR}$ \\
\hline
	&	3.8	&	3.8 \\
\hline\hline
\end{tabular}
\end{center}

This fact, that also appears in the previous Contact Interaction of (VV) and (AA) type
is obvious since in all the three cases there is no parity violation
in the New Physics Lagrangian and therefore the effect in $A_{LR,\mu}$
is depressed as one can easily verify.

\section{Conclusions}
\label{sec:conc}

In this brief communication we have applied the Z peak subtracted
approach to derive bounds on New Physics parameters in several specific models at a
LC with c.m. energy $\sqrt{q^2} = 500\ \mbox{GeV}$ with high
luminosity ${\cal L} = 500\ \mbox{fb}^{-1}$. We have proposed a very
simple description of universal New Physics effects in terms of
three parameters $\delta_z$, $\delta_s$ and $\delta_\gamma$. We
have shown that the longitudinal polarization asymmetry $A_{LR, \mu}$
plays an important role in constraining New Physics, being very
sensitive to the $\delta_s$ parameter. The size of the improvement
depends of course on the particular considered model, but it is
definitely non negligible, reaching a 20\% in the important case of
Anomalous Gauge Couplings. The method can be extended to the analysis of
non universal models; in the specific case of Contact Interactions 
we found significant $15-20\%$ 
improvements of the bounds when $A_{LR, \mu}$ is included in the analysis. 


\end{document}